# Natural configurations and normal frequencies of a vertically suspended, spinning, loaded chain with both extremities pinned

Jean-Marc Noël,[a] Caroline Niquette,[b] S. Lockridge[a] and N. Gauthier[a]

a) Department of Physics, The Royal Military College of Canada, Postal Station Forces 17000, Kingston, K7K 7B4, ON.

b) Department of Physics, McGill University, 3600 rue University, Montréal, Canada, H3A 2T8, QC.

## ABSTRACT

The resonant configurations and normal frequencies of a loaded hanging chain that is rotating uniformly about the vertical are examined from theoretical and experimental perspectives. The chain is assumed pinned at both ends, with an extra load added to the lower end for stability. The equation of motion for this system is solved and it is shown that the various resonant configurations of the chain are described by Bessel functions of order zero. The normal frequencies are obtained and the predictions of the model are compared to experiment. Time-exposure photographs are presented to illustrate the shape of the various normal modes.

## I. Introduction

The dynamical behavior of a massive, vertically hanging chain or cable with a free lower extremity has been studied extensively over the years, in this journal and elsewhere. For example, the early texts of Routh[1] and Lamb[2] discuss the transverse modes of such a system with respect to oscillations that take place in a fixed vertical plane. More recently, in this journal, Bailey[3] discussed the time required by a short transverse pulse to propagate up and down the chain. In other independent studies, Satterly,[4] Levinson,[5] Morse,[6] Coomer et al.[7] and McCreesh *et al.*[8] also considered various aspects of the transverse modes, normal frequencies and swaying of a chain of discrete-links or of a heavy cable. Young[9] considered the longitudinal standing waves on a vertical slinky with a free lower extremity. Other workers studied the effects of rotation around a vertical axis on the stationary configurations that a heavy chain or cable can take, again when the lower extremity is free. Western[10] considered a chain with free lower extremity; he determined its normal frequencies, compared the results to a few experimentally measured modes and also produced time-exposure photographs of the resonant configurations of the system.

Even though the present topic is well established, it is still very relevant today. For example, it has recently been shown, by Allen and Schmidt,[11] for example, that the modes of vibration of a classical, non-relativistic string with a heavy mass at one end are of relevance in the study of quantum strings. Generally speaking, the problem of a heavy loaded chain or cable is also of interest for pedagogical reasons. For one, it can provide an interesting opportunity for students to gain a deeper understanding of the methods of mathematical physics, by working in a concrete manner with Bessel functions. It can further help students to develop a better understanding of more realistic resonating systems, of their deformations, and of their normal modes.

The present article considers the motion of a chain or heavy cable that has its upper extremity fixed to a uniformly rotating vertical axis while its lower extremity is also pinned, yet allowed to rotate about the same vertical axis. The lower extremity of the specimen is further subjected to a non-vanishing external load, for stability. The equation of motion for this system is solved and the resonant configurations and normal modes are obtained and analyzed. As far as we have been able to ascertain, this problem appears not to have been considered previously and we believe that it can provide an interesting undergraduate project or an effective demonstration relating to the normal modes of a more complex system than the idealized cases that are covered at the early level of an undergraduate physics program.

An experimental prototype of the above system was built and the collected data was compared to the predictions of the model.

In Section II of this article, the equation of motion for the system is solved for the normal configurations and modes. Section III describes the experimental set-up that was used to collect data. Section IV analyzes the results and compares them to the predictions of the model. The possible causes of discrepancy between theory and experiment are also examined and commented upon.

### II. The equation of motion, the resonant configurations and normal frequencies

Consider a heavy, uniform inextensible chain (or cable) of total mass $M$ and length $L$ whose position of static equilibrium in the gravitational field is the vertical $z$-axis. The upper extremity of the chain is located at $z = L$ and forced by a motor to rotate around the vertical axis at a known uniform angular velocity $\Omega$. The lower extremity of the chain is located at $z = 0$, where it is pinned and put under an additional tension, $T_0$, through stiff springs. The situation is illustrated in Fig. 1. The chain is coupled to the springs through a bearing that allows the rotation to take place with little friction. The bearing can also move vertically by a very small amount, in order to allow the chain to enter into one of its normal modes of deformation. At resonance, the chain thus moves away from the vertical axis by a horizontal amount $r = r(z)$, which is to be determined. Once a resonant configuration has been established, the boundary conditions are taken to be:

$$r(z = 0) = r(z = L) = 0. \tag{1}$$

In the sequel, $M$ and $L$ will represent the chain's mass and equilibrium length, respectively, with $\lambda = M/L$ the mass per unit length.

The differential equation that was derived by Western[10] for the motion of a rotating vertical chain with a free lower extremity also applies to the present situation, except that the above boundary conditions and the net tension, $T(z)$, are different, where

$$T(z) = T_0 + \frac{Mg}{L} z. \tag{2}$$

In this expression, $g$ is the acceleration due to gravity and $T_0$ is the additional tension that is applied at the bottom of the chain. The form of the resulting equation for the stationary solutions of the motion is given by:

$$\frac{\partial}{\partial z}\left[T(z)\frac{\partial}{\partial z}r(z)\right] + \Omega^2 \lambda r(z) = 0. \tag{3}$$

In order to solve this equation for the allowed angular frequencies, $\Omega$, and the corresponding configuration functions, $r(z)$, we first set

$$x = \frac{z}{L} + \frac{T_0}{Mg}. \tag{4}$$

Inserting Eq. (4) into Eqs. (2) and (3) then gives, with $\sigma^2 \equiv \Omega^2/g$, that:

$$\frac{\partial^2 r}{\partial x^2} + \frac{1}{x}\frac{\partial r}{\partial x} + \sigma^2 \frac{r}{x} = 0. \tag{5}$$

Finally, introduce a second change of variable, thus:

$$w^2 = 4\sigma^2 x. \tag{6}$$

Equation (5) then becomes:

$$\frac{\partial^2 r}{\partial w^2} + \frac{1}{w}\frac{\partial r}{\partial w} + r = 0. \tag{7}$$

This is a Bessel equation of the first kind and of order zero whose general solution is given by:

$$r(z) = C_1 J_0\left(\frac{2\Omega}{g}\sqrt{\frac{1}{\lambda}(\lambda g z + T_0)}\right) + C_2 Y_0\left(\frac{2\Omega}{g}\sqrt{\frac{1}{\lambda}(\lambda g z + T_0)}\right). \tag{8}$$

In this expression, $J_0$ and $Y_0$ are Bessel functions of the first kind, of order zero;[12] also, $C_1$ and $C_2$ are constants to be determined by imposing the boundary conditions of Eq. (1). The resulting system of two homogeneous equations then has a nontrivial solution only if

$$J_0(\gamma\Omega) Y_0\left(\gamma\Omega\sqrt{\frac{\lambda g}{T_0}+1}\right) - Y_0(\gamma\Omega) J_0\left(\gamma\Omega\sqrt{\frac{\lambda g}{T_0}+1}\right) = 0. \tag{9}$$

The following variable was introduced, for simplicity:

$$\gamma = \frac{2}{g}\sqrt{\frac{T_0}{\lambda}}. \tag{10}$$

The discrete normal frequencies of the motion, $\Omega = \Omega_m$, with $m = 0,1,2,...$, are obtained by solving Eq. (9) numerically; this is done in Section IV, where a comparison between theory and experiment is made.

We now turn our attention to the description of the experimental set-up that was used to collect the data.

### III. Experimental set-up and the collected data

The experimental set-up consisted of a steel lab equipment rack, a variable DC power supply, a permanent magnet motor, a wooden block, two fish weighing scales and a stroboscope. The motor was securely bolted to the top of the rack while a wooden block was attached to the bottom, through a bearing that allowed rotation around the vertical axis. The wooden block was permitted to move in a vertical direction but two fish weighing scales restricted its upward motion thereby providing additional tension. The appropriate specimen, a chain or cable, was attached to the motor and wooden block using eyebolts and spring link snaps. We used spring link snaps to attach the chain and/or cable so the specimen could be changed easily. Finally, the apparatus was anchored to the floor to increase the stability.

A variable DC power supply was used to control the angular velocity of the motor and consequently, the angular velocity of the chain (or cable). The chain rotated freely about its vertical axis and after some time, attained a stable configuration. The angular velocity of the chain was measured using the stroboscope while the applied tension was measured using the fish weighing scales. The measured values for the angular velocities and tensions are tabulated in Tables 1 through 3 for two chains and a cable that were used in this experiment. The same wooden block was used for all the experiments; its mass of 0.714 kg exerted a constant tension of 6.997 N at the lower extremity of the specimen. Three specimens were examined: a brass chain, a stainless steel chain and a steel cable.

In Figure 1, we present a photograph of the apparatus while one of the chains used is rotating in its stable third mode. The motor is at the top of the rack while the block and fish scales are at the bottom, just outside the field of view of the camera. The spring link snaps can be seen at the top and bottom extremities of the chain.

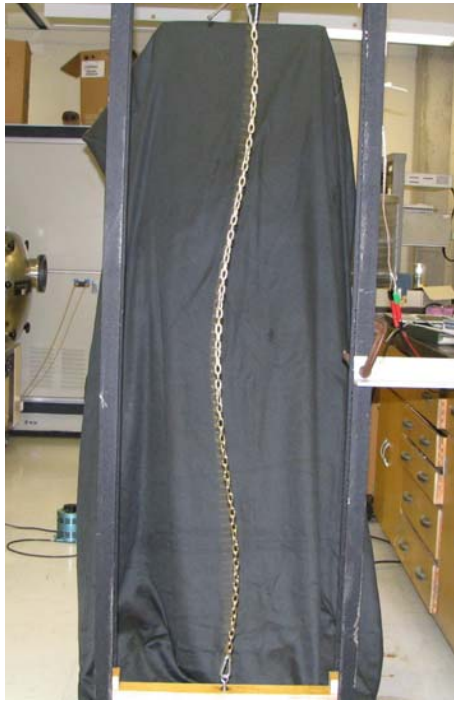

**Figure 1- A photograph of the apparatus in action taken with a digital camera. The chain is shown rotating in its stable third mode (second overtone).**

Table 1 contains the experimental results for the brass chain. The fundamental and third modes were both observed using this chain. The fundamental has one anti-node and two nodes (one at each end). The third mode, or second overtone, has three anti-nodes and four nodes. The experimental results of Table 1 are presented in two parts: the upper and lower parts relate to the fundamental mode and the second overtone, respectively. The measured angular velocities are tabulated in rpm in the first column and in rad/sec in the second column. The third column gives the tensions read from the fish weighing scales and the fourth column contains the total tension applied to the lower extremity of the chain. The results for the steel cable are summarized in Table 2 using the same format as in Table 1. Likewise, the results for a stainless steel chain are summarized in Table 3.

During our experiments, only the brass chain was able to rotate in its second overtone for a period long enough for us to determine the tensions in the fish scales as well as the angular velocity. The other chain and cable were unstable in that mode and would quickly begin to rotate in their fundamental modes.

**Table 1** - Observed angular velocities and tensions using the brass chain. The mass of the chain was 0.277 kg and has a length of 1.360 m.

| Mode 0 (Fundamental) | | | |
|---|---|---|---|
| $\Omega$ (rpm) | $\Omega$ (rad/s) | $T_{scales}$ (N) | $T_0$ (N) |
| 423 | 44,29 | 78,4 | 85,39 |
| 494 | 51,73 | 117,6 | 124,59 |
| 632 | 66,18 | 205,8 | 212,79 |
| 647 | 67,75 | 215,6 | 222,59 |
| 649 | 67,96 | 215,6 | 222,59 |
| Mode 2 (second overtone) | | | |
| $\Omega$ (rpm) | $\Omega$ (rad/s) | $T_{scales}$ (N) | $T_0$ (N) |
| 700 | 73,30 | 19,6 | 26,60 |
| 775 | 81,15 | 39,2 | 46,20 |
| 840 | 87,96 | 34,3 | 41,30 |
| 896 | 93,82 | 19,6 | 26,60 |
| 960 | 100,53 | 44,1 | 51,10 |
| 992 | 103,88 | 24,5 | 31,50 |
| 1000 | 104,71 | 58,8 | 65,80 |
| 1068 | 111,84 | 29,4 | 36,40 |
| 1100 | 115,19 | 39,2 | 46,20 |
| 1170 | 122,52 | 78,4 | 85,40 |
| 1180 | 123,56 | 39,2 | 46,20 |
| 1292 | 135,29 | 58,8 | 65,80 |
| 1421 | 148,80 | 98,0 | 105,00 |
| 2144 | 224,51 | 68,6 | 75,60 |
| 2272 | 237,92 | 73,5 | 80,50 |

**Table 2** – Observed angular velocity and tension using a steel cable. The mass of the cable was 0.091 kg and its length was 1.310 m. The format is the same as Table 1.

| Mode 1 (Fondamental) | | | |
|---|---|---|---|
| $\Omega$ (rpm) | $\Omega$ (rad/s) | $T_{scales}$ (N) | $T_0$ (N) |
| 511 | 53,51 | 19,6 | 26,60 |
| 647 | 67,75 | 39,2 | 46,20 |
| 753 | 78,85 | 58,8 | 65,80 |
| 858 | 89,84 | 78,4 | 85,40 |
| 1009 | 105,66 | 107,8 | 114,80 |
| 1111 | 116,34 | 137,2 | 144,20 |

Table 3 - Observed angular frequencies and tension for the stainless steel chain. The mass of the chain was 0.699 kg and its length was 1.352 m.

| Mode 1 (Fundamental) | | | |
|---|---|---|---|
| $\Omega$ (rpm) | $\Omega$ (rad/s) | $T_{scales}$ (N) | $T_0$ (N) |
| 303 | 31,73 | 117,6 | 124,60 |
| 343 | 35,91 | 137,2 | 144,20 |
| 345 | 36,12 | 137,2 | 144,20 |
| 369 | 38,64 | 166,6 | 173,60 |
| 394 | 41,25 | 196,0 | 203,00 |
| 446 | 46,70 | 313,6 | 320,60 |
| 462 | 48,38 | 254,8 | 261,80 |
| 490 | 51,31 | 274,4 | 281,40 |
| 521 | 54,55 | 333,2 | 340,20 |
| 536 | 56,12 | 352,8 | 359,80 |
| 558 | 58,43 | 372,4 | 379,40 |
| 575 | 60,21 | 411,6 | 418,60 |
| 581 | 60,84 | 392,0 | 399,00 |

## IV.    Analysis and discussion

Figure 2 presents a graph of the numerical solutions of equation (9); the first three roots are given. The fundamental mode has the lowest frequency and is shown as the lower curve, labeled *m=0*. The first overtone is the next lowest frequency and the corresponding curve has been labeled *m=1*. Finally the second overtone has been labeled *m=2*.

The vertical axis of Figure 2 corresponds to $\gamma\Omega$, where $\Omega$ is the observed angular velocity of the chain or cable and the horizontal axis corresponds to the total tension that was applied at the bottom of the chain. We note from each of the theoretical curves that as the tension increases the resonant angular frequencies also increase; this is in line with expectations. The observed frequencies and tensions for all three specimens used in the experiment have been superimposed for comparison. The results are denoted as follows: stainless steel chain (inverted triangles); brass chain (asterisks); cable (circles). The associated error bars have also been included. Notice that the observed data falls very close to the theoretical curve for the mode corresponding to *m=0*. The results for the second overtone exhibit a larger deviation from the predicted behavior. However when the error bars for the data are taken into consideration, the general trend of the data agrees reasonably well with that predicted by the model used.

Among the reasons why the experimental results differ from those predicted by the model, the most important is perhaps the cork-screw-like motion that the chain undergoes while rotating. Indeed, each link of the chain must rotate through a small angle before the next link can be carried along and made to rotate also. The corresponding transfer of angular motion down the chain then causes the chain to rotate in a cork-screw-like manner rather than in a vertical plane and friction in the bearing at the lower end would likely force the chain to maintain that twist in steady state. This particular aspect of the motion was not built into the model used. A secondary source of uncertainty can be associated with the fish weighing scales that we used to apply the tension at the bottom of the chain. Using better-quality and well-calibrated scales would likely result in a decrease in size of the horizontal component of the error bars.

The theoretical shape of the chain was also compared to the actual physical shape of the chain. Fig. 3 shows an example of a frame taken from a movie clip showing the physical shape of the two chains while they were rotating in their fundamental modes. The left hand panel shows the numerical results that were obtained from our model, using Eq. 8. The middle and right-hand panels show frames for the stainless steel and the brass chains, respectively. One can see from these picture-frames that the general features of the predicted shapes are in agreement with the actual shapes of the rotating specimens.

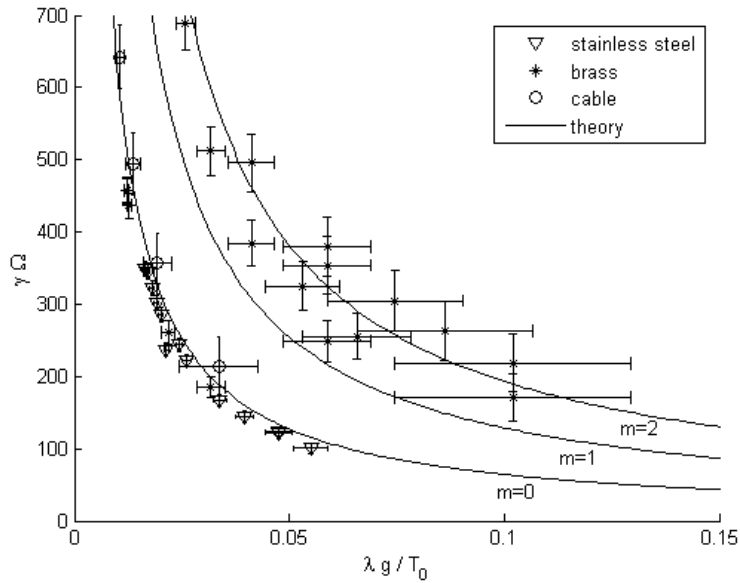

**Figure 2 - The numerical solutions of equation (9) for the first 3 modes (lines). The observed frequencies are plotted for the stainless steel chain (inverted triangles), the brass chain (asterisks) and cable (circled). Also shown are the error bars associated to each observation.**

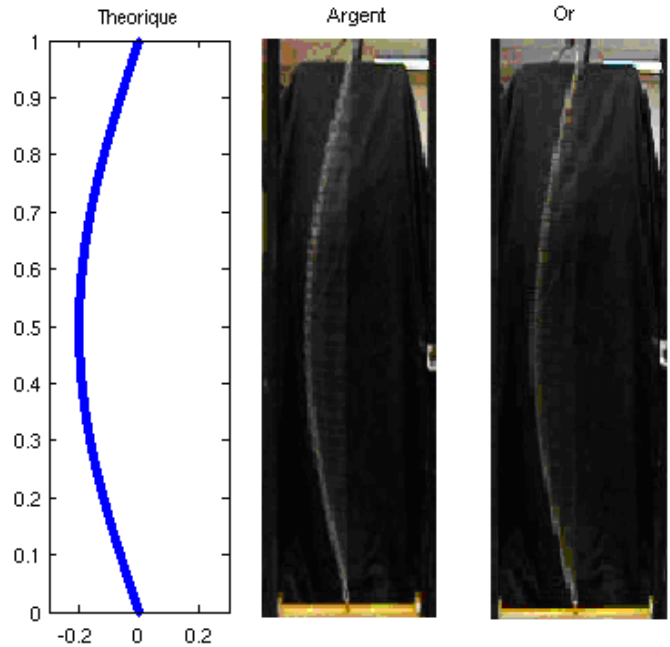

**Figure 3 - The shape of the rotating chain. The theoretical result is shown in the left panel: the stainless steel and brass chains are shown in the middle and in the right panels, respectively.**

## IV. Concluding remarks

The motion of a vertical chain that is pinned at both ends and made to rotate around the vertical in the gravitational field was considered. A simple model for the rotational dynamics of the system was used and the resulting equation of motion solved for the stationary configurations and normal frequencies. The predictions of the model were compared to data collected from three different specimens, two chains and one cable. The results for the lower modes agree reasonably well with the predictions of the model. Friction at the lower bearing of the assembly appears to have been the main source of deviation between the predictions of the model and the experimental results.

We suggest that a study of the above system can have pedagogical value by allowing students to develop a better understanding of an application of Bessel functions and of the complex behavior of a real system.